\documentclass{IEEE-Con-Sys-mag}
\usepackage{amsmath,amssymb,amsfonts}
\usepackage{algorithmic}
\usepackage{graphicx}
\usepackage{textcomp}
\usepackage[compress]{cite}

\usepackage{subcaption,epsfig,gensymb,hyperref}

\jvol{}
\jnum{}
\paper{}
\jmonth{}
\jname{Internal Report}
\pubyear{}

\begin{document}
\title{Teaching data-driven control 
	\stitle{from linear design to adaptive control with throttle valves}}
\author{EMMANUEL WITRANT, IOAN DOR\'E LANDAU, and MARIE-PIERRE VAILLANT, \today}

\maketitle
	\dois{}{}

\begin{summary}
Electric throttle valves represent a challenge for control design, as their dynamics involve strong nonlinearities, characterized by an asymmetric hysteresis.
Carrying experiments on multiple valves, a large variability in the characteristics of each valve and erratic steady-state behaviors can also be noticed, impairing classical model-based control strategies.
Nevertheless, local data-driven linear models can be obtained and simple proportional-integral (PI) controllers, tuned individually for each valve with the appropriate data set, provide good tracking performance.
As these controllers cannot be transposed from one valve to another, a robust strategy and an adaptive controller (using identification in closed-loop and controller re-design) may be necessary to propose a general method.
This work aims at promoting control education on a simple yet challenging process, going from frequency analysis and linear design to an adaptive control method implemented with an online recursive algorithm.
\end{summary}


\chapterinitial{E}{lectric} throttle valves are the most frequently devices used in industry for flow control.
The valve considered in this paper is more specifically a butterfly  valve, which regulates the downstream pressure by adjusting the rotation of a disk.
Such valves have a relatively low cost and a fast time response, and are classically used in automotive, chemical, pharmaceutical and food industries.
Nevertheless, the electro-mechanical apparatus of throttle valves induces complex dynamics that need to be handled with care to satisfy precise flow regulation objectives.

Throttle valve control is a particularly challenging topic, which motivated dedicated research as soon as in the 1960s, when an output feedback design is proposed to obtain a high accuracy steam valve control \cite{Johnson1965}. 
The process is modelled as a second order system and a frequency analysis leads to the design of a proportional feedback with reasonable gain.
Considering the valve regulation in a large steam turbine-generator unit, the authors in \cite{Callan1965} conclude that a lead-lag compensator is more suitable, and that PI control may result in limit cycles when combined with the valve dead-zone for inadequate gain values.
The complexity of the environment is considered one step further by  \cite{Kwatny1975}, where the impact of acoustic waves in the piping of a boiling unit is considered. The concepts of state variable feedback and dynamic observer are used in this case to control a simplified model.

More recently, electric throttle valves triggered the interest of many researchers in automatic control, as the static friction effects and the nonlinearities of the gear box and of the return spring induce particularly complex dynamics (usually modeled with an asymmetric hysteresis).
A brief review of some works supported by experimental evaluation is given here.
Linear approaches are proposed, for example, with linear quadratic control \cite{Cassidy1980}, robust $\mathcal{H}_\infty$ design \cite{Vargas2014}, and linear parameter-varying (LPV) modeling and mixed constrained H$_2$/H$_\infty$ control  \cite{Zhang2015}.
The valve friction has motivated specific control strategies, such as the one proposed by the authors of \cite{Canudas2001}, who derived a dynamic model including friction and aerodynamic torque, and proposed an adaptive pulse control strategy. 
A hybrid feedforward–feedback friction compensator with friction parameter adaptation is proposed in \cite{PANZANI201342} and an adaptive PID feedback controller with adaptive feedforward compensators for friction, limp-home and backlash is proposed in \cite{Jiao2014}.
Some other methods focus on handling the asymmetry of the return-spring: it is taken into account with a nonlinear asymmetric PI controller in \cite{Pujol2016} and, combined with the friction effect, with a hybrid LPV method in \cite{hamze:tel-02524432}.
Sliding mode control methods also brought some interesting contributions, with a robust adaptive chattering-free strategy using a genetic algorithm in \cite{Mao2020} and an adaptive scheme based on recursive terminal sliding mode in \cite{Hu2021}.

Simple experiments designed to evaluate control methods on throttle valves using Arduino\textsuperscript{\tiny\textregistered} are described in \cite{Martins2012}, which compares analog and digital implementations of a PI controller. The improvement of proportional-derivative and proportional-integral-derivative (PID) over proportional controllers is noticed in \cite{Supriyo2015} and in \cite{Jadhav2016}, which includes the valve (PID-controlled) in a drive-by-wire setup.
Throttle valves are used as an educational theme in \cite{Lucena2013}, to motivate the learning of microprocessors systems.

The objective of this work is to use the ubiquity of electric throttle valves in an Arduino\textsuperscript{\tiny\textregistered} set up to teach both some fundamental aspects of automatic control and some more advanced methods.
Two classes, given at University Grenoble Alpes at the Master level, motivated (and now rely on) this work.
The first class is on \emph{Modeling and System Identification}, mostly based on the textbooks \cite{Lju99,Landau2006}.
The throttle valve lab associated with this class follows the philosophy of the pioneering works (pre-1980) cited above: focusing primarily on the observation of an appropriated data set, a simple linear model is obtained for each valve and a PI digital controller is implemented.
The second class is on \emph{Adaptive Control}, following the digital control approach proposed in \cite{Landau11}.
The lab starts by noticing that a controller designed for one valve cannot be easily applied to another, due to the model variability.
This fact motivates an adaptive method to handle the system complexity, as it is done in most of the recent research results.
The aim of this paper is to describe some key aspects of the labs in a pedagogical way such that the throttle valves can be easily used for teaching purposes and possibly adapted to other classes.

The paper is organized as follows. 
The steady-state and transient behaviors of the valve are described first, along with linear modeling and PI control design.
An adaptive control method based on closed-loop output error identification and controller redesign is addressed in then proposed.
Specific highlights are given on the throttle valve experiments, on robust control design and its usage in a digital control setup, and on the organisation of the teaching labs.

\begin{sidebar}{Experimental test bench of the throttle valve}

\setcounter{sfigure}{0}
\renewcommand{\thesfigure}{S\arabic{sfigure}}

\sdbarinitial{T}he block diagram description and the electronic setup of the experimental test bench of the throttle valve are depicted in Figures~\ref{Fig:Throttle_pic} and \ref{Fig:Throttle_electronics}, respectively.
The valve has the commercial reference 03L128063, a valve that equipped some Audi cars before 2010 and has only one spring (some other car engine throttle valves often considered in the literature are more complex, with a non-zero limp home position that results in a system with two hystereses).
This rotational spring is set on the shaft of the valve plate, exerting a torque that counteracts the motor's torque thus resulting in a control of the angular position of the plate (torsional energy storage by the spring, which integrates the velocity).
The opening angle is thus regulated by modulating the input voltage (driving the motor torque).
This is done by generating a piece-wise modulation (PWM, ranging from 0\% to 100\% when the input signal is modulated from 0 to 255) of the voltage with an Arduino Mega 2560\textsuperscript{\tiny\textregistered} board, programmed with the \emph{Arduino Integrated Development Environment}\textsuperscript{\tiny\textregistered} (IDE) software.

\sdbarfig{\epsfig{file=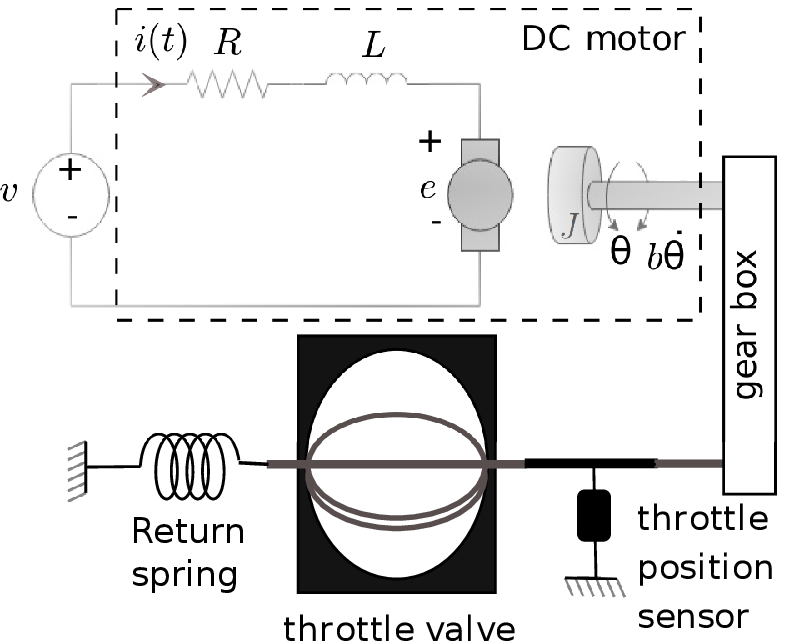,width=.8\linewidth,clip=}}{Block diagram of the electric throttle valve.\label{Fig:Throttle_pic}}

The valve is equipped with a programmable magnetic angle sensor KMA221 that provides an analog output ratio-metric to the supply voltage (set to 5\,V to match the specification of Arduino\textsuperscript{\tiny\textregistered}'s analog inputs).
The analog-to-digital converter of the Arduino\textsuperscript{\tiny\textregistered} board then converts this signal into a 10-bits one when using the command \verb"analogRead()".

The 12\,V DC motor of the valve is controlled using the SHIELD-MD10 board, a Cytron\textsuperscript{\tiny\textregistered} 10A motor driver shield for Arduino\textsuperscript{\tiny\textregistered}.
This shield uses an NMOS H-Bridge to achieve a speed control PWM frequency up to 10\,kHz.
The default PWM frequency is increased to avoid an annoying high-frequency noise generated by the valve (the free-running timer 3 of ATmega32 is modified using \verb"TCCR3B").
The motor is controlled through two digital pins of the Arduino\textsuperscript{\tiny\textregistered} board (selected by the mini jumpers on the shield): one for direction (\verb"pinDIR", set to \verb"HIGH" or \verb"LOW") and one for the velocity (\verb"pinPWM", between 0 and 255).

\sdbarfig{\epsfig{file=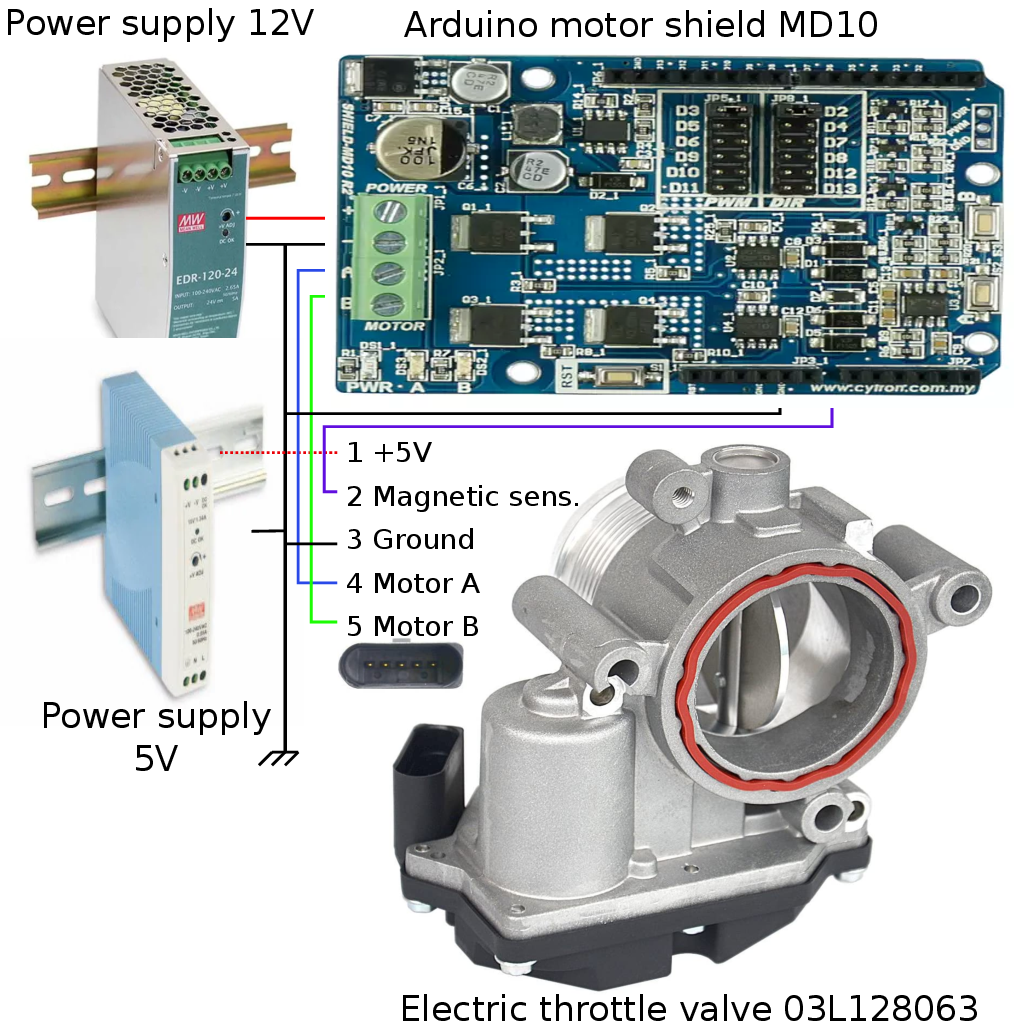,width=1\linewidth,clip=}}{Experimental test bench: power supply and wiring of the electric throttle valve.\label{Fig:Throttle_electronics}}

A similar experimental test bench was initially developed as a research topic for car industry \cite{hamze:tel-02524432} and we simplified it for teaching purposes.
The initial throttle valve used for testing the adequacy with teaching linear control is referred to as \emph{Experiment 0} in the sequel.
Once satisfactory results have been obtained,  10 other valves have been made available for building a control lab teaching equipment (7 of them are considered in the paper for comparison purposes).
While we ordered valves with the same commercial reference as \emph{Experiment 0}, it was not the same seller and not the same series.
This has a serious impact on the experiments, because of important dispersion of the valves static and dynamic characteristics.
The proposed strategy is thus to develop the algorithms for \emph{Experiment 0}, which has a more predictable behavior, and test their efficiency on the other valves.

The cost of an experimental test bench with the components shown in Figure~\ref{Fig:Throttle_electronics} is about 140~\texteuro \, (it could be reduced by choosing a simpler Arduino\textsuperscript{\tiny\textregistered} board). 
This results in a low-cost test bench that is easy to connect.
Another advantage is the robustness of an industrial process that has to ensure a reasonable lifetime in the (unfriendly) car engine environment. 
\end{sidebar}

\section{From non-linear dynamics to linear control} \label{Sec:Modeling}

The experimental test bench of the throttle valve described in the sidebar provides interesting data sets to investigate nonlinear dynamics and varying time-constants.
Furthermore, major differences can be noticed when comparing the responses of different valves.
Theses complexities are analyzed in this section and frequency analysis shows that a linear behavior can still be captured and used to design a simple PI feedback controller.

\begin{pullquote}
	``frequency analysis shows that a linear behavior can still be captured and used to design a simple PI feedback controller''
\end{pullquote}

\subsection{Steady-state behavior and time constants}

\begin{figure}[!th]%
	\centering
	\epsfig{file=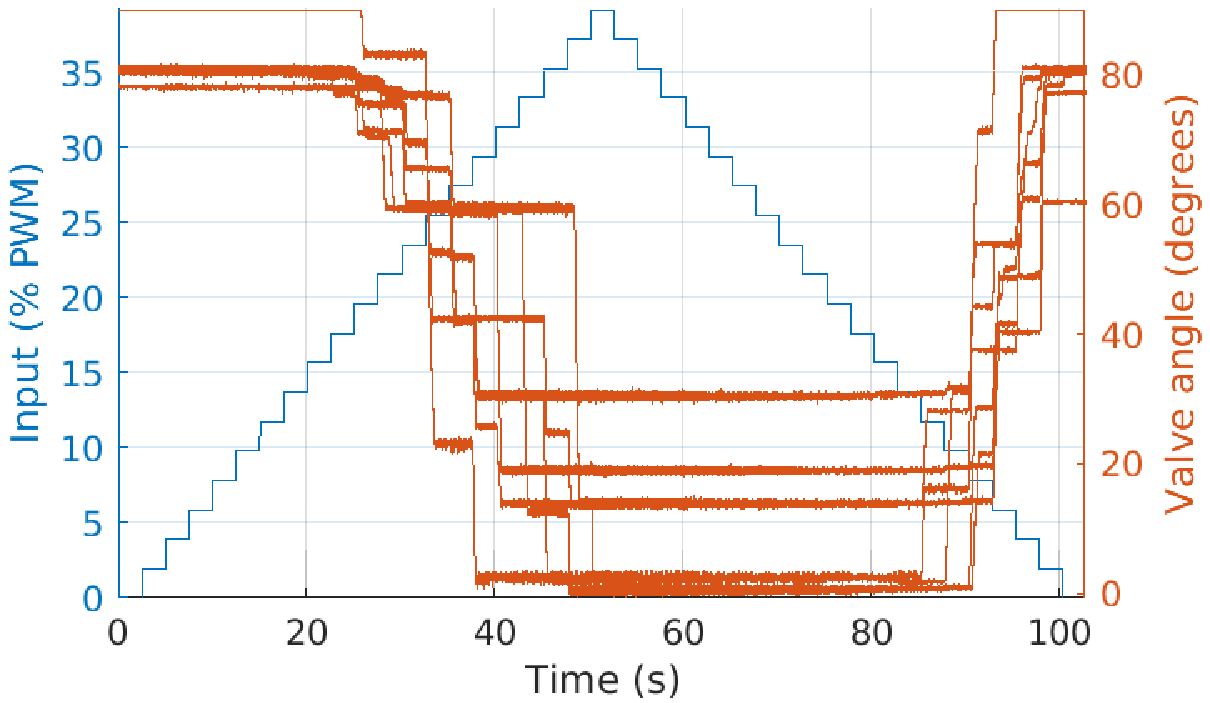,width=1\linewidth,clip=}
	\caption{Time response of the height different valves to a sequence of steps. Left axis: input PWM voltage (blue curve). Right axis: opening angles of the eight valves (orange curves). A large variability is observed on the static responses of the valves.}
	\label{Fig:ident_preliminary_gain_superposed}
\end{figure}

The steady-state behavior of the valves is investigated by applying a sequence of steps to the input voltages (increases and decreases of the PWM signals by 5). 
Each step is maintained during 2.5\,s, to ensure that each experimental test bench has enough time to reach the steady-state value. The various results are superposed on Figure\,\ref{Fig:ident_preliminary_gain_superposed}, which shows the superposed responses of eight different valves.
The starting times and magnitudes of the different hystereses vary significantly.
The measured angle corresponding to the fully open position (when the PWM is set to 0) is 90\degree \, for \emph{Experiment 0} (used for calibration) and 80\degree \, for the others, showing some calibration discrepancies.
We can also note that the angle reached at the maximum PWM input (40\,\% here) differs from valve to valve, probably due to different friction coefficients.

Repeating the sequence of input steps multiple times on a given valve, a large variability of the hysteresis shape and of the angles corresponding to the open or close positions can be noticed.
The use of a nonlinear model based on a hysteresis, as it is classically done in the modern automatic control literature, should thus be done carefully as the hysteresis parameters vary largely from valve to valve and even during time for a given valve.

The response time of the valves is evaluated by zooming on the first second following an input increase or decrease and also has a large variability from valve to valve and for a given valve.
The rise time $t_R$ (to go from 10\% to 90\% of the final values) ranges from 0.2 to 0.8\,s, approximately.
The sampling time is chosen following the guidelines proposed in \cite{Landau2006} (two to nine samples per rise time) as $T_s$ = 50\,ms. 

\subsection{Frequency-domain analysis}

The first step to model the process from a data set is to generate a signal that is sufficiently rich in terms of frequency content.
This can be done efficiently, for example, by using a Pseudorandom Binary Sequence (PRBS) \cite{Lju99,Landau2006} obtained from $N_r$ shift registers with feedback.
The PRBS design is performed here according to the method proposed in \cite{Landau2006}.
As the longest step of the PRBS has to encompass the longest rise time and as we wish to limit the PRBS length, we use the constraint
$$
p N_r T_s > t_R,
$$
where $p$ is the ratio between the PRBS frequency and the sampling frequency.
Choosing $p=2$ sets $N_r = 9$ and a PRBS of length $2^{N_r}-1=511$, which is reasonable to implement as a pre-computed table in the Arduino\textsuperscript{\tiny\textregistered} board (computed during the call of the \verb|setup| function, prior to the \verb|loop| function).

\begin{figure}[!th]%
	\centering
	\epsfig{file=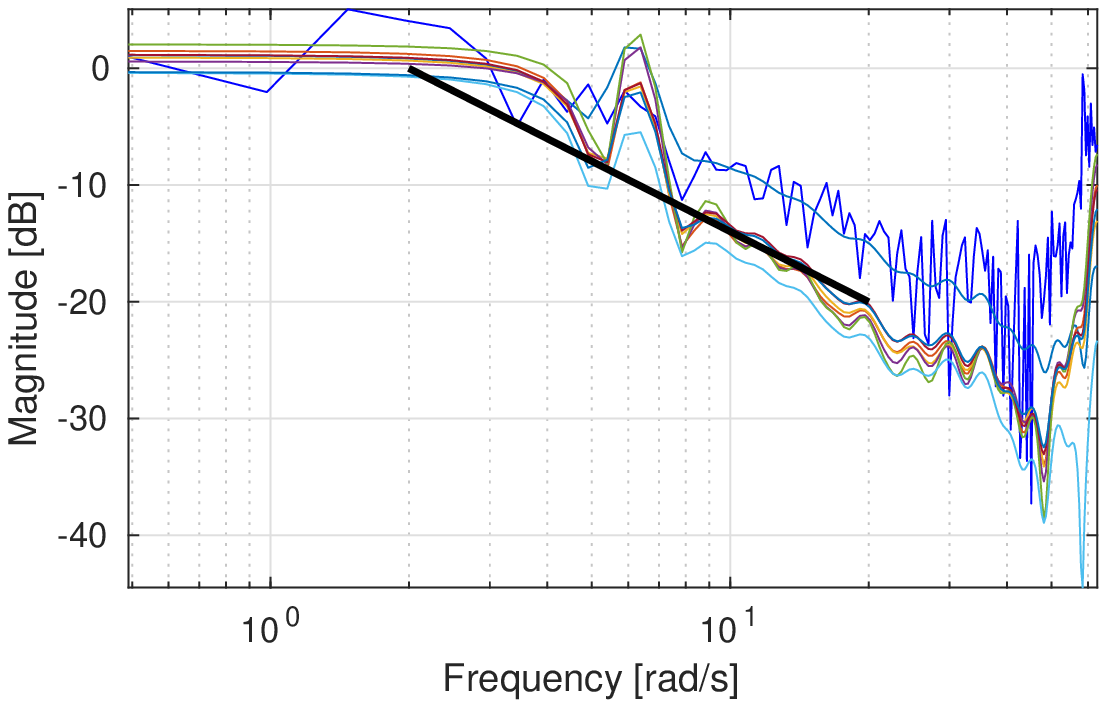,width=1\linewidth,clip=}
	\caption{Empirical Transfer Function Estimate (ETFE) of the valves responses to a PRBS input. The raw ETFE of \emph{Experiment 0} is depicted in dark blue while filtered values are shown in the other colors for all the valves. The black line follows a slope of -20dB/dec, which captures the main transient dynamics of the valves.}
	\label{Fig:ident_prbs_ETFE}
\end{figure}

The PRBS is used to generate a PWM input signal centered at 16\%, modulated by an amplitude between 10\% and 14\% that depends on the valve (the larger magnitude overcomes the Coulomb friction for all the valves).
The frequency response of each valve is obtained by calculating the Empirical Transfer Function Estimate (ETFE, defined as the ratio between the discrete Fourier transforms of the output and of the input sequences \cite{LJUNG19851653}) for each experiment using Matlab\textsuperscript{\tiny\textregistered}.
The raw ETFE of \emph{Experiment 0}, the smoothed ETFEs (by a Hann window of size 25, as suggested by \cite{Lju99}) of all the experimental test benches, and the slope of -20~dB/dec are depicted in Figure\,\ref{Fig:ident_prbs_ETFE}. 
We can note the following:
\begin{itemize}
	\item the smoothed ETFE provides a reasonable approximation of the ETFE below 45~rad/s;
	\item the ETFEs of the different experiments have a similar behavior, except for \emph{Experiment 0} that has a larger gain;
	\item a slope of -20~dB/dec reasonably approximates the ETFE slope, motivating the use of models with a single pole;
	\item the change of slope at 3\,rad/s suggests a time constant of 0.5\,s (consistent with the observed rise time).
\end{itemize}

\subsection{Linear models} \label{sec:LinMod}

We consider a class of auto-regressive models with exogenous inputs that  writes as
\begin{eqnarray}
	y(t) &=& - a_1 y(t-1) - \ldots - a_{n_a} y(t-n_a) \nonumber \\
	&+ & b_1 u(t-1) + \ldots + b_{n_b}u(t-n_b) = \phi^T(t)\theta, \label{eq:linMod}
\end{eqnarray}
where $y(t)$ is the valve angle, $u(t)$ is the PWM input, $n_a$ and $n_b$ define the number of past samples (outputs and inputs, respectively) used to compute the actual output, and $\{a_1,\, \ldots,\,a_{n_a}, \, b_1,\, \ldots,\,b_{n_b} \}$ are constant parameters that form the \emph{parameter vector}
\begin{eqnarray}
	\theta &=& [a_1, \, \ldots, \,a_{n_a}, \, b_1, \, \ldots, \,b_{n_b}] . \label{eq:theta}
\end{eqnarray}
The past data necessary to compute the value at $t$ is stored in the regressor (measurement vector)
\begin{eqnarray}
	\phi(t) &=& [-y(t-1),\,-y(t-2),\,\ldots, -y(t-n_a), \nonumber \\
	& & \quad u(t-1),\,\ldots, u(t-n_b)]^T . \label{eq:phi}
\end{eqnarray}
The dimensions of the variables are $u(t),y(t)\in \mathbb{R}^1$, $\theta, \phi \in \mathbb{R}^n$, $n=n_a+n_b$, where $\mathbb{R}^n$ is the real n-dimensional Euclidean space.

The data sets used to estimate the  model parameters are generated by a PRBS.
The predicted output at time $t$ using the data set available at time $t-1$ is
\begin{eqnarray*}
	\hat{y}(t|\hat{\theta}) &=& \phi^T(t)\hat{\theta},
\end{eqnarray*}
where the vector containing the estimated parameters is
\begin{eqnarray*}
	\hat{\theta} &=& [\hat{a}_1, \, \ldots, \,\hat{a}_{n_a}, \, \hat{b}_1, \, \ldots, \,\hat{b}_{n_b}].
\end{eqnarray*}
The optimal set of parameters is obtained by minimizing the least-squares criterion
\begin{eqnarray}
	V_{{N}}(\hat{\theta},Z^{{N}}) &=& \frac{1}{\bar{N}}\sum_{t=\bar{n}+1}^N \frac{1}{2} \left( y(t)-\phi^T(t)\hat{\theta} \right)^2, \label{eq:LinModV}
\end{eqnarray}
where $\bar{n} = \max\{n_a,n_b\}$, $\bar{N} = N - \bar{n}$, and $Z^{N}$ represents the data set that contains the inputs and outputs for $t=1,\ldots,\, N$ used for estimation.

\begin{figure}[!th]%
	\centering
	\epsfig{file=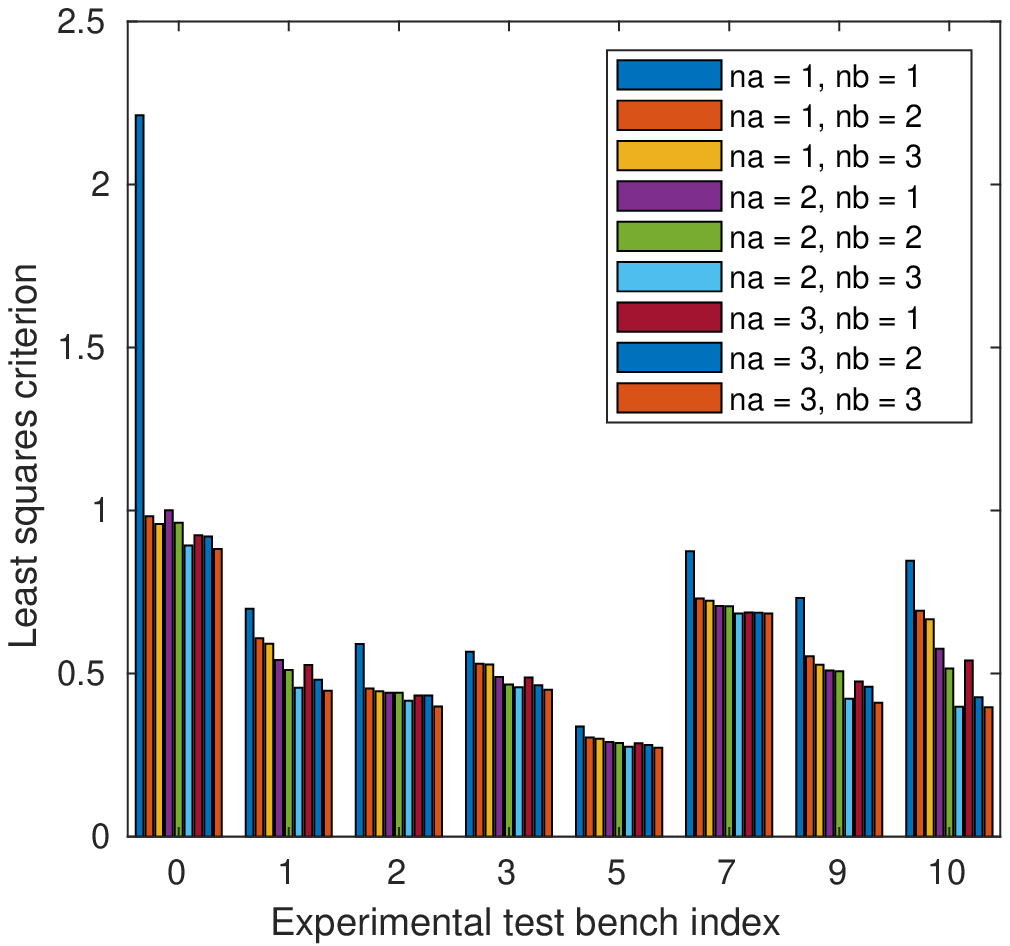,width=1\linewidth,clip=}
	\caption{Normalized least-squares criterion \eqref{eq:LinModV} for different combinations of the number of free parameters of the model (varying both the number of past outputs $n_a$ and past inputs $n_b$ used for the predicted output). The same computations are carried for each valve and presented in parallel.
	While increasing the number of parameters improves the accuracy, the model obtained with $n_a=n_b =1$ gives reasonable results for most of the valves.}
	\label{Fig:ident_prbs_V}
\end{figure}

Computing the criterion $V$ for $n_a$ and $n_b$ varying between 1 and 3, we obtain the results presented in Figure~\ref{Fig:ident_prbs_V}.
Note that similar results (qualitatively) were obtained using Akaike's information criterion, a classical criterion to assess the suitable model order.
The results on Figure~\ref{Fig:ident_prbs_V} show that, for most valves, having three free parameters ($n_a=1$ and $n_b=2$, or $n_a=2$ and $n_b=1$) decreases $V$ in comparison with the $n_a=n_b =1$ case.
Nevertheless, the model with $n_a=n_b =1$ still gives reasonable results except for \emph{Experiment 0} (for which the least-squares criterion is doubled).
Increasing the order of the denominator $n_a$ (globally for all $n_b$) does not significantly decreases $V$, as expected from the ETFE analysis.
Some valves (such as \emph{Experiment 10}) can get more benefits from a more complex model than others (such as \emph{Experiment 5} or \emph{7}).

To be consistent with the frequency analysis and to simplify the control design, we consider only linear models characterized by $n_a=1$ and $n_b=1$ for the controller design.

\subsection{PI feedback control} \label{sec:PIdesign}

As a reference feedback control, we consider the digital implementation of a PI control that fullfills a pole placement objective on the close-loop system.
The digital control proposed by \cite{Landau2006} can then be used directly as
\begin{equation}\label{eq:PI}
u(t) = u(t-1) - r_0 y(t) - r_1 y(t-1) + (r_0 + r_1) r(t) ,
\end{equation}
where $r(t)$ is the desired reference and $r_0$ and $r_1$ are the controller's gains.
Introducing the unit delay operator $q^{-1}$ such that $y(t-1)=q^{-1}y(t)$, the gains are computed using the identified model parameters and set the desired denominator of the discrete closed-loop transfer function to $1 + p_1 q^{-1} + p_2 q^{-2}$ with
\begin{equation}\label{eq:PIparam}
r_0 = \frac{p_1 - \hat{a}_1 + 1}{\hat{b}_1} \quad \textrm{and} \quad r_1 = \frac{p_2 + \hat{a}_1}{\hat{b}_1}.
\end{equation}
A pole placement design specifying $p_1$ and $p_2$ can thus be directly implemented in the controller using \eqref{eq:PI}-\eqref{eq:PIparam}.
The closed-loop poles associated with the parameters $p_1$ and $p_2$ result from the choice of second order dynamics having specified damping and time response (obtained for example using the diagrams in \cite{Landau11}).

\begin{figure}[!th]%
	\centering
				\begin{subfigure}{0.45\textwidth}
	\epsfig{file=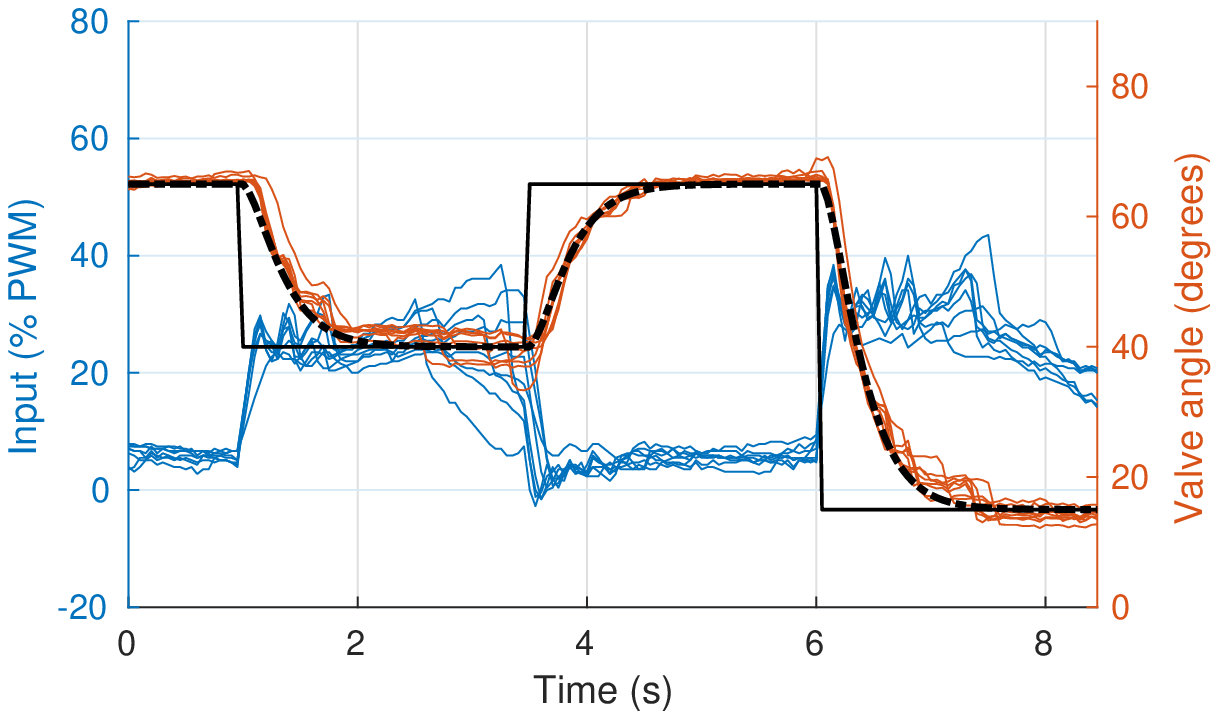,width=1\textwidth,clip=}
				\caption{Damping ratio = 1.0, rise time = 0.8 s.}
			\end{subfigure}
			\begin{subfigure}{0.45\textwidth}
	\epsfig{file=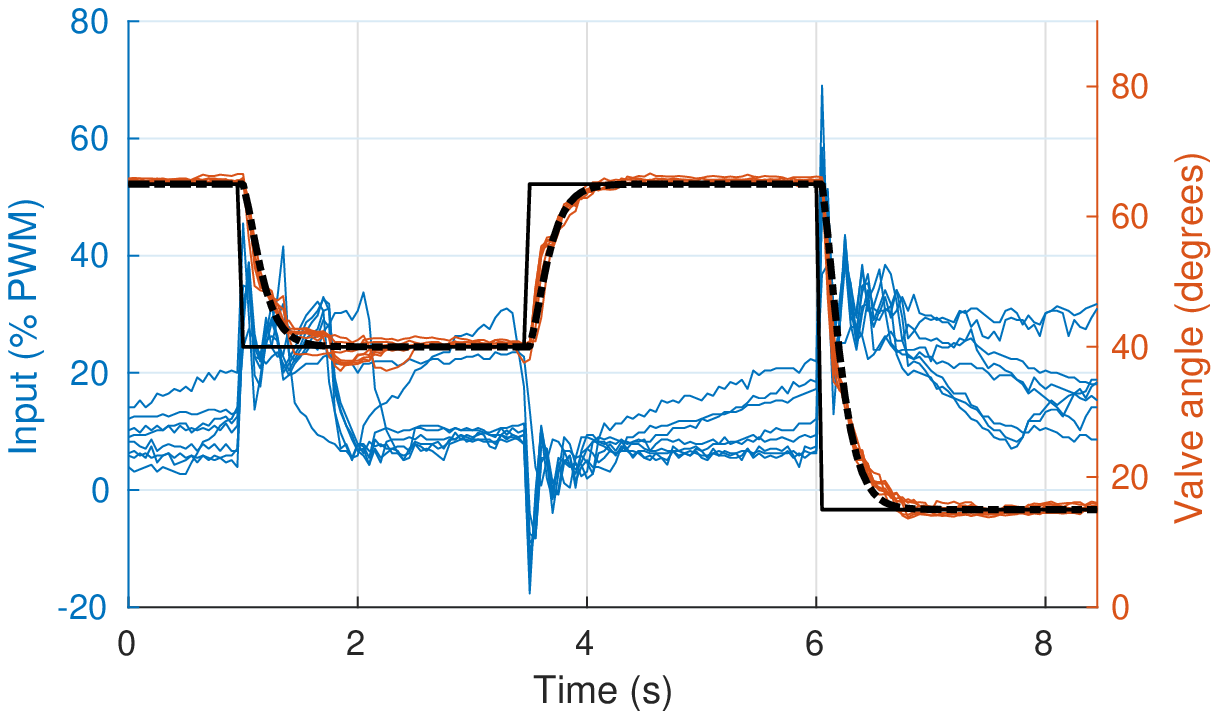,width=1\textwidth,clip=}
				\caption{Damping ratio = 1.0, rise time = 0.4 s.}
			\end{subfigure}
			\caption{Closed-loop responses of the valves with PI feedback control: superposed (for the 8 experimental test benches). Left axis: PWM inputs (blue). Right axis: closed-loop responses (orange), tracked reference (continuous black line) and expected output (dash-dot black line).
			Both control designs follow the expected closed-loop trajectories closely.}
	\label{Fig:pi_2designs}
\end{figure}

This controller is implemented on the experimental test benches as follows.
First, trial and error testing on a test bench led us to the conclusion that a target second order system that is critically damped (that is, with a damping ration $\zeta = 1$) provides good results.
Two different response times are considered: one with a target $t_R$ = 0.8\,s and a more demanding control with $t_R$ = 0.4\,s. 
For each valve, $r_0$ and $r_1$ are computed using the specific values of $a_1$ and $b_1$ identified from the PRBS response of the corresponding valve.
The controller's efficiency to track a reference is investigated for $r(t) =  40 \pm 25\,\degree$, thus over a large operating range.
The results are depicted on Figure~\ref{Fig:pi_2designs}, where the outputs are also compared with the dynamics expected from the closed-loop denominator $1 + p_1 q^{-1} + p_2 q^{-2}$.
We can observe that the closed-loop responses are particularly consistent between the valves, despite the previously discussed differences and nonlinearities.
The responses also follow the expected closed-loop responses closely, specially for the higher-gain design ($t_R$ = 0.4\,s).

\begin{sidebar}{Robust control design} 

\setcounter{sequation}{0}
\renewcommand{\thesequation}{S\arabic{sequation}}

\setcounter{sfigure}{2}
\renewcommand{\thesfigure}{S\arabic{sfigure}}

The objective of the PI feedback control \eqref{eq:PI}-\eqref{eq:PIparam} is to set the simplest digital PI controller for the throttle valve. It starts with the hypothesis that a first order discrete-time model with $n_b=n_a=1$ can be identified from data.
It is however very important to discuss the robustness of the design with respect to neglected dynamics and variations of the plant parameters.
In fact (even in continuous time) the first order model is a rough approximation of reality. 
In addition, some high-frequency dynamics (also called ``parasitic'' dynamics) are always present and should be taken into account in the identification and the design stage (if this aspect may be neglected in the identification stage, hypotheses should be made upon the existence of this neglected dynamics in order to take it into account in the design stage). 
This high-frequency dynamics can often be modelled as an additional fractional delay which will lead to a model with $n_b=2$ (see Sections~7.5.1-7.5.2 in \cite{Landau2006} for details). 
Indeed, the previous identification results have shown that a better model validation is obtained using $n_b=2$: this effectively indicates the presence of a fractional delay.

\sdbarfig{\includegraphics[width=0.9\columnwidth]{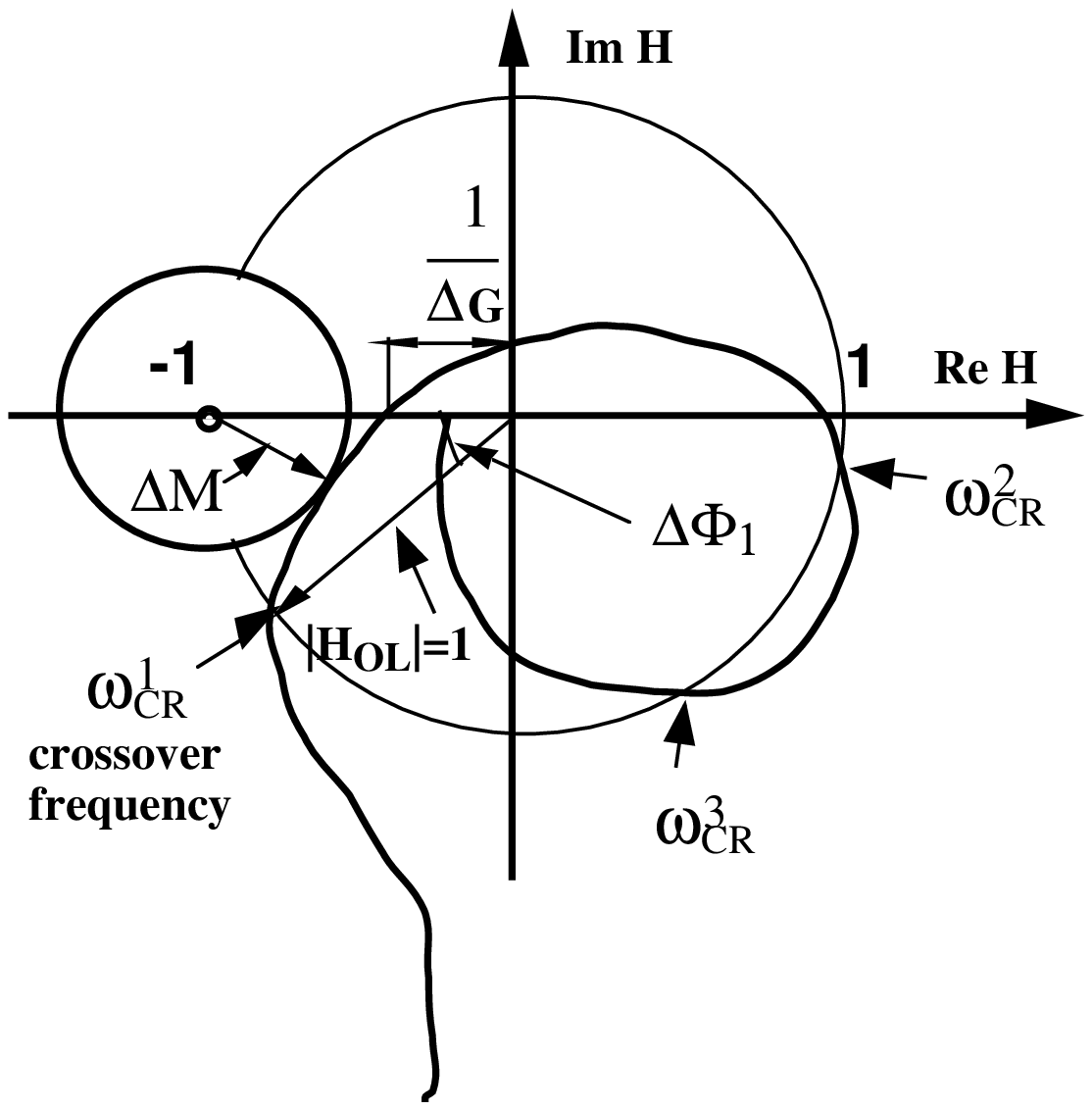}}{Robustness indicators on a Nyquist plot.\label{fig7_5}}

Independently of considering a more accurate model, the robustness of the design with respect to plant model uncertainties (neglected high-frequency dynamics, variations of the plant parameters) should be evaluated. The robustness of the design is assessed by examining the sensitivity functions in the frequency domain. 
For the specific problem considered we focus on two sensitivity functions: 1) the output sensitivity function, 2) the input sensitivity function.
First note that the linear dynamics described by \eqref{eq:linMod} write in the standard transfer operator form \cite{Landau2006}
\begin{equation}
\frac{y(t)}{u(t)} = \frac{b_1 q^{-1} + \ldots + b_{n_b} q^{-n_b}}{1 + a_1 q^{-1} + \ldots + a_{n_a} q^{-n_a}} = \frac{B(q^{-1})}{A(q^{-1})}. \label{eq:TFbase}
\end{equation}
The digital controller canonical structure is considered with the RST formulation as
\begin{equation}
S(q^{-1}) u(t) = - R(q^{-1}) y(t) + T(q^{-1}) r(t). \label{eq:RST}
\end{equation}
Note that the discrete-time transfer functions are expressed in a complex frequency-domain ($z-domain$) by replacing $q^{-1}$ with $z^{-1}$.
With these notations, the \textit{output sensitivity} function is defined as:
\begin{eqnarray*}
S_{yp}(z^{-1}) & = & \frac{A(z^{-1})S(z^{-1})}{A(z^{-1})S(z^{-1})+B(z^{-1})R(z^{-1})}, \\
 & = & \frac{A(z^{-1})S(z^{-1})}{P(z^{-1})},
\end{eqnarray*}
where P defines the computed poles of the closed loop. 
The \textit{input sensitivity} function is
\begin{eqnarray*}
S_{up}(z^{-1}) &=& \frac{-A(z^{-1})R(z^{-1})}{A(z^{-1})S(z^{-1})+B(z^{-1})R(z^{-1})}, \\ 
&=& \frac{-A(z^{-1})R(z^{-1})}{P(z^{-1})}.
\end{eqnarray*}

The first robustness indicator is the ``modulus margin'' $\Delta M$. Figure~\ref{fig7_5} allows defining this quantity as the radius of the circle centered in the Nyquist point $[-1,j0]$ and tangent to the Nyquist frequency plot of the open loop transfer function $H_{OL}(z^{-1})$, where
\begin{equation*}
H_{OL}(z^{-1})=\frac{B(z^{-1})R(z^{-1})}{A(z^{-1})S(z^{-1})}.
\end{equation*}
It is the minimum distance between the instability point and the Nyquist plot of the open loop transfer function.  Inspection of Figure~\ref{fig7_5} leads to the relationship
\begin{eqnarray*}
\Delta M &=&|1+H_{OL}(j\omega)|_{min} = |S_{yp}^{-1}(j\omega)|_{\min}, \\
&=&(|S_{yp}(j\omega)|_{\max})^{-1}.
\end{eqnarray*}
If we consider the graph of the magnitude of $S_{yp}$ in the frequency domain, we have that (in dB)
\begin{equation}
|S_{yp}(j\omega)|_{max} = \Delta M^{-1} = -\Delta M .
\end{equation}
The aim of the design is to ensure a modulus margin $\Delta M \geq 0.5=-6dB$.
The controller should thus be designed such that
$|S_{yp}(j\omega)|_{max} \leq 6 dB$.
To satisfy this condition for the simple PI control design, the dominant dynamics (dominant poles) have to be chosen by selecting a second order system with a certain damping and resonance frequency and/or to increase the value of an auxiliary pole.

The \textit{input sensitivity} function characterizes the tolerance with respect to neglected dynamics and parameters variations (particularly in the high-frequency range). It can be shown \cite{Landau2006}  that for guaranteeing the closed loop stability, the tolerated additive uncertainty should satisfy the condition
\begin{equation}
|\frac{B'}{A'}-\frac{B}{A}|< |S_{up}^{-1}(j\omega)|,
\end{equation}
where $\frac{B'}{A'}$ defines the perturbed plant model and $\frac{B}{A}$ is the plant model used for control design. 
This equation can be interpreted as follows: a good tolerance to additive plant uncertainties is obtained at the frequencies where $|S_{up}(j\omega)|$ is small, and conversely a low tolerance to additive plant model uncertainties occurs at the frequencies where $|S_{up}(j\omega)|$ has a large value. 
Since the uncertainties are mainly located in the high frequencies range, it is desirable to get the lowest possible value for $|S_{up}(j\omega)|$  in this frequency range.
\end{sidebar}

\begin{sidebar}{}
\setcounter{sfigure}{3}
\renewcommand{\thesfigure}{S\arabic{sfigure}}	

Let us consider the design using the plant model with $a_1= -0.9152$ and $b_1=-0.0609$ (parameters identified for \emph{Experiment~0}).
The desired dominant poles are defined by a second order system with $\omega_0=5$ and a damping $\zeta=1$ (resulting from a desired time response of 0.8\,s). The corresponding controller parameters are: $r_0=-5.8719$ and $r_1=5.0685$.  
Figure \ref{SRobustness} shows the magnitude Bode plot of the output sensitivity function. The maximum of $|S_{yp}|$ is less than $6 dB$: the modulus margin is thus larger than $0.5$. Therefore the design is satisfactory from the point of view of the minimum distance with respect to the Nyquist point.

\sdbarfig{\includegraphics[width=\columnwidth]{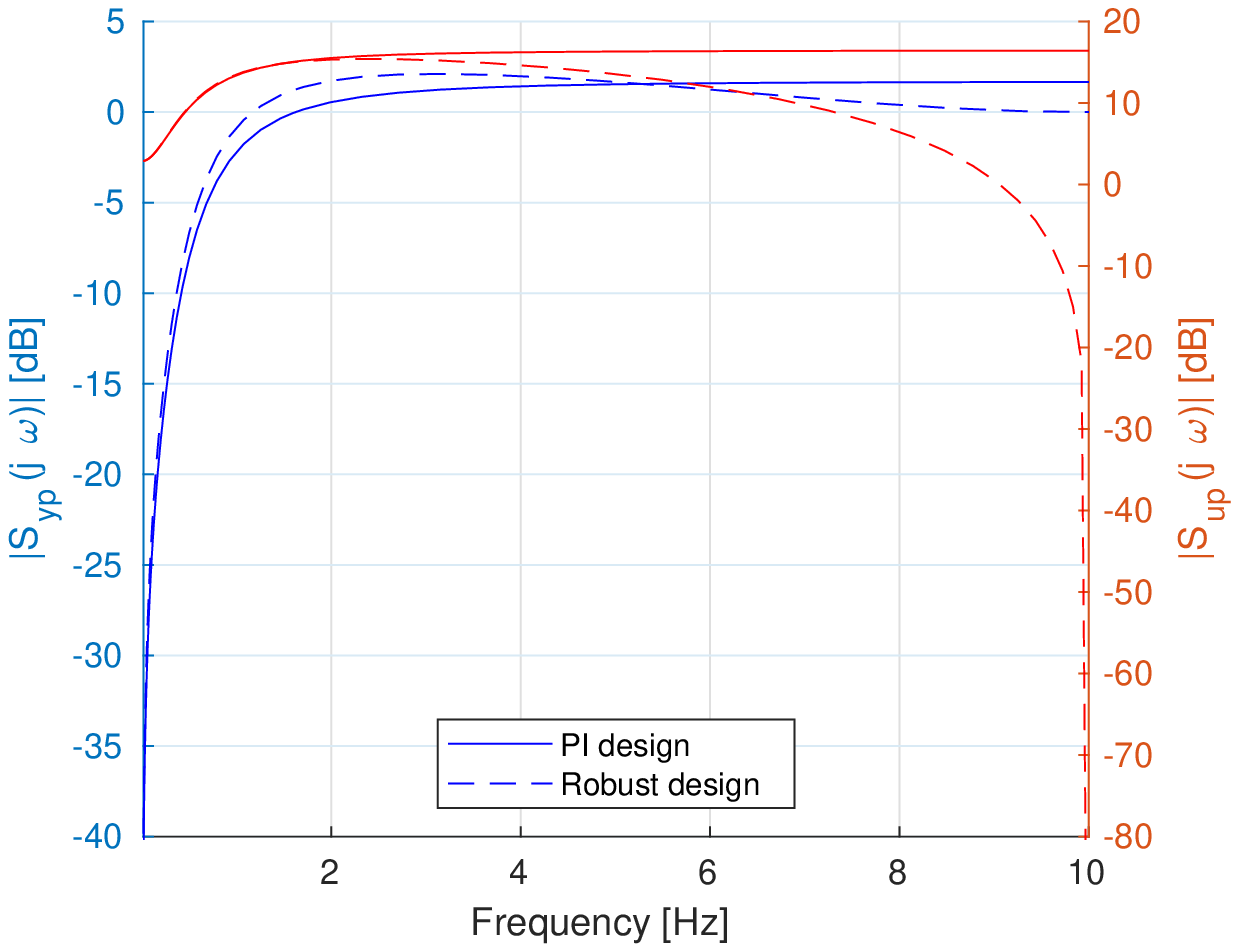}}{Sensitivity functions for the robustness analysis ($a_1=0.9152; b_1=-0.0609$). Left axis (blue): output sensitivity; right axis (orange): input sensitivity. Plain lines: PI design setting the closed-loop behavior as a damped 2$^{nd}$ order dynamics; dashed lines: robust design when opening the loop at $0.5\,f_s$. High frequency disturbances are removed from the control signal by the robust design. \label{SRobustness}}
	
Figure \ref{SRobustness} also shows the magnitude bode plot of the input sensitivity function. 
The value of $|S_{up}(j\omega)|$ is large in the high frequencies regions, which implies a low tolerance of the design with respect of the neglected high-frequency dynamics. 
An improved design should lower as much as possible $|S_{up}(j\omega)|$ in the high frequencies region (equivalently the controller should have a very low gain in high frequencies). 
Unfortunately, with the actual complexity of the controller, we do not have enough degrees of freedom to achieve both performance and low value of the sensitivity $|S_{up}(j\omega)|$ at high frequencies. 
To achieve this, a fixed part is added in the $R$ polynomial of the controller
\begin{equation*}
R(q^{-1}) = R' (q^{-1}) (1+q^{-1}),
\end{equation*}
where $R' (q^{-1})$ is the new polynomial that needs to be designed (to achieve the pole placement objective).
The term $1+q^{-1}$ has a zero gain at $0.5\, f_s$ ($f_s$ being the sampling frequency) and the input sensitivity is thus zero at this frequency.
Similarly, the integral effect is included with the contraint
\begin{equation*}
S(q^{-1}) = S'(q^{-1}) (1-q^{-1}),
\end{equation*}
where $1-q^{-1}$ sets the integral action and $S' (q^{-1})$ has to be designed.
	
To design this controller we have to solve a Bezout equation of the form
\begin{equation*}
AS'(1-q^{-1})+BR'(1+q^{-1})=P=P_D P_F,
\end{equation*}
for given $A$, $B$ and $P$ (decomposed into dominant poles $P_D$ and auxiliary poles  $P_F$ that can be added to improve robustness), and the unknowns $R'$ and $S'$. The resulting controller has the RST form \eqref{eq:RST}.

Specifically for the previous example if we open the loop and do not change the assigned dominant poles $P_D$ ($P_F=1$), the coefficients of the resulting controller are: $r_0= -3.0157$, $r_1=-0.4017$, $r_2=2.6140$, and $s_0= 1.0000$, $s_1=-0.8261$, $s_2=-0.1739$. The corresponding frequency characteristics of the output and input sensitivity functions are shown in Figure~\ref{SRobustness} (dashed line). In comparison with the previous PI design (continuous line), the modulus of the new input sensitivity function goes towards $0$ ($- \infty$~dB) at high frequencies close to $0.5f_s$, which is not the case when the loop is not opened at $0.5\,f_s$ (10\,Hz). The influence on the output sensitivity function is minor (it reaches 0\,dB at 10\,Hz, which indicates that the system will be in open loop at this frequency). 

It is interesting to compare now the two controllers in terms of performance. 
Simulated step responses of the two closed-loop systems (not shown) are indistinguishable: ideally, both controllers achieve the same desired pole placement.
The controllers are evaluated on another experiment (\emph{Experiment 6}) with the idea that, from an industrial perspective, one would want the controller to perform equally well on all the devices of the same brand. 
The experimental results of a tracking scenario are shown in Figure \ref{Fig:Robust_perf}. 
The robust controller mostly achieves a better tracking, whether close to the linearization setpoint (40$^\circ$) or far from it (references at 70$^\circ$ and 10$^\circ$).

\sdbarfig{\includegraphics[width=\columnwidth]{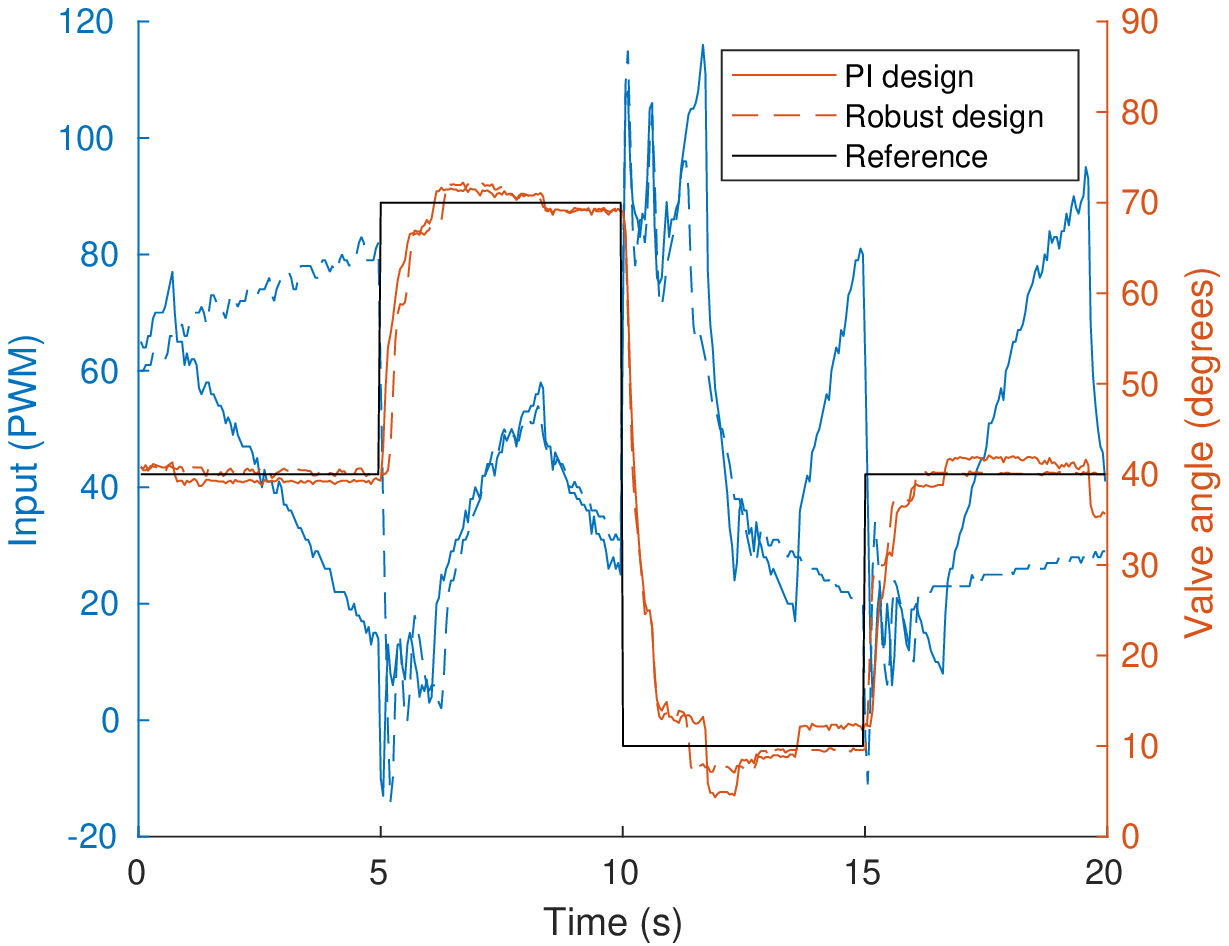}}{Evaluation of the robust design performance on \emph{Experiment 6}.
			Left axis: PWM inputs in blue, right axis: closed-loop responses in orange and tracked reference in black.
			 Plain lines: PI design setting the closed-loop behavior as a damped 2$^{nd}$ order dynamics; dashed lines: robust design when opening the loop at $0.5\,f_s$. Robustness does not impair the controller's performance. \label{Fig:Robust_perf}}

\end{sidebar}

\section{Adaptive Control} \label{Sec:Adaptive}

The final objective of data-driven control is to use data acquired in closed loop operation to improve the performance of the closed loop. 
This approach should also be able to take into account the possible variations of the plant parameters during operation, which may cause serious performance degradation.

The set of techniques dedicated to the data-driven tuning of the controllers is called "Adaptive Control". 
Adaptive control can be viewed as an add-on loop on top of the current feedback loop, whose objective is to achieve and to maintain a desired performance for the closed loop. 
\begin{pullquote}
	 ``an adaptive control system is able to estimate in real-time the parameters of the plant model and re-tune in real-time the controller''
\end{pullquote}
The basic idea is that an adaptive control system is able to estimate in real-time the parameters of the plant model and re-tune in real-time the parameters of the controller based on the current estimated plant parameters (without the designer in the loop!). 
If these two operations, parameter estimation and controller re-design, are done at each sampling time, one has a genuine adaptive control system. 
If a time separation is introduced between these two operations (the adaptation system identifies the plant during a certain time horizon in the presence of a fixed controller and then based on these estimated parameters re-designs the controller and applies it), one has the technique called \textit{iterative identification in closed loop and controller re-design} (implemented at the end of this section on the experimental test bench).
However, to initialize the procedure in the absence of an initial controller, an identification in open loop operation is necessary, followed by the design of the controller based on the identified model.

Two aspects have to be taken into account.\\
1) Since the system is operated in real-time and we would like to have an estimation of the plant model in real-time (as the plant evolves in time), one has to move from off-line identification (which uses a set of input/output data) to on-line identification, which updates the parameters of the plant model at each sampling instant. 
To implement an on-line identification procedure one should use recursive algorithms for plant model estimation.\\
2) For the design of the first controller, identification of the plant model has to be done in open loop operation. However, when doing \textit{iterative identification in closed loop and controller re-design}, the identification is done in closed loop. In this context there is a feedback from the plant output to the plant input via the controller and this alters the performance of open loop identification algorithms. In closed loop operation the identification paradigm (objective) is different from the open loop case. 
As a consequence the recursive parameter estimation used for identification in open loop should be modified.

We start by reviewing recursive algorithms for on-line identification in open loop and then present the modified algorithms dedicated to identification in closed loop.

\subsection{Recursive open loop identification algorithms} 
To formalize the problem, consider the discrete-time model  described by \eqref{eq:linMod} to \eqref{eq:phi}.
An adjustable prediction model is defined as
\begin{equation}\hat{y}^0(t+1)=\hat{y}[(t+1) | \hat{\theta}(t)]=
\hat{\theta}^T(t) \phi(t),
\label{eq:Prediction}
\end{equation}
where $\hat{y}^0(t+1)$ is termed the \emph{a priori} predicted output, as it depends upon the vector of estimated parameters at instant $t$
\begin{equation}
\hat{\theta}(t)= [\hat{a}_1(t),\, \ldots,\,\hat{a}_{n_a}(t), \, \hat{b}_1(t),\, \ldots,\, \hat{b}_{n_b}(t)]^T.
\label{eq:Estpar}
\end{equation}
It is very useful to consider also the \emph{a posteriori} predicted output, computed on the basis of the new estimated parameter vector at $t+1$, $\hat{\theta} (t+1)$, which will be 
available somewhere between $t+1$ and $t+2$. 
The \emph{a posteriori} predicted output\index{a posteriori predicted output} is
\begin{align}
\hat{y}(t+1)&=\hat{y}[(t+1) | \hat{\theta}(t+1)] =\hat{\theta}^T(t+1) \phi(t) .
\label{a postout}
\end{align}
Using these predictions, an \emph{a priori} prediction error
\begin{equation}
\epsilon^0(t+1)=y(t+1)-\hat{y}^0(t+1)
\label{eq:apriori}
\end{equation}
and an \emph{a posteriori} prediction error
\begin{equation}
\epsilon(t+1)=y(t+1)-\hat{y}(t+1)= [\theta - \hat{\theta}(t+1)]^T\phi(t)
\label{eq:aposteriori}
\end{equation}
are defined.

The objective is to find a recursive parameter adaptation algorithm with memory. The structure of such an algorithm is
\begin{align} 
\hat{\theta}(t+1) & =\hat{\theta}(t)+\Delta \hat{\theta}(t+1) \nonumber\\
&= \hat{\theta}(t) + f[\hat{\theta}(t), \phi(t), \epsilon^0(t+1)].
\label{eq:standard}
\end{align}
The correction term $f[\hat{\theta}(t), \phi(t), \epsilon^0(t+1)]$ must depend solely on the 
information available at the instant $(t+1)$ when $y(t+1) $
is acquired (last measurement $y(t+1)$, $\hat{\theta} (t)$, and a finite amount of information at 
times $t$, $t-1$, $t-2$, $\cdots$, $t-n$).
An off-line (non-recursive) identification algorithm has been used above with the objective of minimizing the least squares  criterion given in \eqref{eq:LinModV}. 
The recursive version of the \textit{least squares identification} algorithm is (see \cite{Landau2006,Landau11} for details) is
\begin{align}
& \hat{\theta}(t+1) = \hat{\theta}(t) + F(t) \phi(t) \epsilon(t+1), \label{eq_3_2_51} \\
& F(t+1)^{-1} = F(t)^{-1} + \phi(t) \phi^T(t) \label{eq_3_2_52}, \\
& \Leftrightarrow F(t+1) = F(t) - \frac{F(t) \phi(t) \phi^T(t)F(t)}{1+\phi^T(t)F(t)\phi(t)}, \label{eq_3_2_53}\\
& \epsilon(t+1) = \frac{y(t+1) - \hat{\theta}^T(t) \phi(t)}{1+\phi^T(t)F(t)\phi(t)}. \label{eq_3_2_54}
\end{align}
In practice, the algorithm is started  at $t = 0$ by choosing
\begin{equation} \label{eq_3_2_55}
F(0)=\frac{1}{\delta}I=(GI)I;~0<\delta\ll 1,
\end{equation}
where $I$ is the identity matrix of appropriate dimensions and $GI$ is the initial gain (typical value $GI=1000$). 
The recursive least squares algorithm is an algorithm with a decreasing adaptation gain.
As an example, consider the estimation of a single parameter. 
In this case, $F(t)$ and $\phi(t)$ are scalars, and \eqref{eq_3_2_53} becomes
\begin{displaymath}
F(t+1) = \frac{F(t)}{1+\phi(t)^2F(t)} \le F(t); \qquad \phi(t),~F(t) \in R^1.
\end{displaymath}
The recursive least squares algorithm thus gives less and less weight to the new prediction errors and to the new measurements.
This type of variation of the adaptation gain is not suitable for the estimation of time-varying parameters, and other variation profiles for the adaptation gain must be considered. 

In order to estimate time-varying parameters, the recursive formula for the inverse of the adaptation gain $F\left(t+1\right)^{-1}$ given by \eqref{eq_3_2_52} is generalized by introducing two weighting sequences $\lambda_1(t)$ and $\lambda_2(t)$ that satisfy the constraints
\begin{equation*}
\begin{split}
F(t+1)^{-1} = \lambda_1(t)F(t)^{-1} + \lambda_2(t)\phi(t)\phi^T(t),  \\
0<\lambda_1(t)\le 1 \;,  \; 0\le \lambda_2(t)<2 \;,  \; F(0)>0 .
\end{split}
\end{equation*}
Note that $\lambda_1(t)$ and $\lambda_2(t)$ have opposite effects.
$\lambda_1(t)<1$ tends to increase the adaptation gain (the gain inverse decreases) while $\lambda_2(t)
>0$ tends to decrease the adaptation gain (the gain inverse increases).
For each choice of sequences $\lambda_1(t)$ and $\lambda_2(t)$ corresponds a \emph{variation	profile} of the adaptation gain and an interpretation can be inferred in terms of the error criterion (which is minimized by the parameter adaptation algorithm).

Using the \emph{matrix inversion lemma} (see \cite{Landau2006,Landau11} for details) gives the adaptation gain evolution
\begin{equation*}
F(t+1) = \frac{1}{\lambda_1(t)} \left[ F(t) - \frac{F(t)\phi(t)\phi^T(t)F(t)}
{\frac{\lambda_1(t)}{\lambda_2(t)} + \phi^T(t)F(t)\phi(t)} \right] .
\end{equation*}
One of the most used option, both for identification of systems with constant parameters or with slowly time varying parameters, is the so-called ``variable forgetting factor''.
In this case
\begin{equation*} \lambda_2(t) = \lambda_2 =1  \end{equation*}
and the forgetting factor is
\begin{equation} \label{eq_3_2_71}
\lambda_1(t) = \lambda_0 \lambda_1(t-1)+1-\lambda_0 \; ,  \; 0<\lambda_0<1 .
\end{equation}
The typical ranges for the initial and constant terms are $\lambda_1(0) = 0.95 \; \mbox{to} \; 0.99$, $\lambda_0=0.5 \;\mbox{to} \;0.99$ (some frequently used values are: $\lambda_1(0) = \lambda_0=0.97$).
$\lambda_1(t)$ can be interpreted as the output of a first order filter $\left(1-\lambda_0\right)/\left(1-\lambda_0q^{-1}\right)$ with a unitary steady state gain and an initial condition $\lambda_1(0)$.
Relation~\eqref{eq_3_2_71} leads to a forgetting factor that asymptotically converges towards $1$ (decreasing adaptation gain).
This type of profile, when used for the model identification of stationary systems, avoids a too rapid decrease of the adaptation gain, thus usually resulting in an acceleration of the convergence  (by maintaining a high gain at the beginning when the estimates are far from the optimum values).

\subsection{Recursive identification in closed loop operation} \label{sec:CLOE}

\begin{figure}[ht]
	\begin{center}
		\includegraphics[width=\columnwidth]{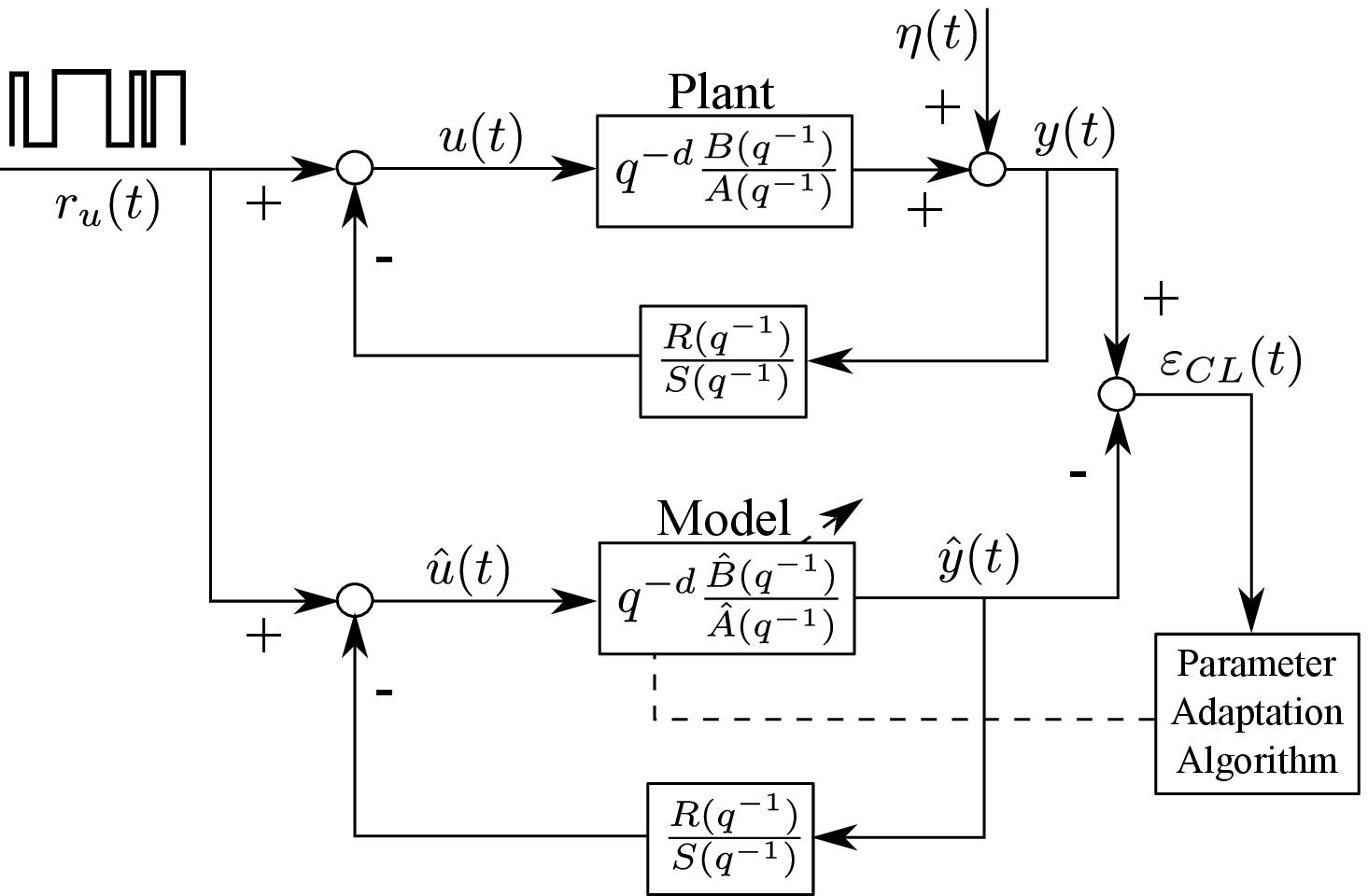}
		\caption{Identification in closed loop. The excitation signal is superposed to control output and the model parameters are adapted to minimize the closed-loop prediction error.}
		\label{fig8_2a}
	\end{center}
\end{figure}

Using an open loop recursive identification algorithm in closed loop operation, in the presence of a fixed controller (situation which is encountered in the \emph{iterative identification in closed loop and controller re-design} method), often does not provide a reliable model of the plant because one identifies the plant in closed loop with the controller.
The identification paradigm calls for the identification of the plant model that gives the best prediction of the closed loop output for a given controller. 
The principle of closed-loop output error identification algorithms is illustrated in Figure~\ref{fig8_2a}.  The upper part represents the true closed-loop system and the lower part represents an adjustable predictor of the closed-loop. This closed-loop predictor uses the same controller as the one used on the real-time system.

The prediction error between the output of the real-time closed-loop system and the closed-loop predictor (closed-loop output error) is a measure of the difference between the true plant model and the estimated one. 
This error can be used to adapt the estimated plant model such that the closed-loop prediction error is minimized (in the sense of a certain criterion). In other words, the objective of the identification in closed-loop is to find the best plant model which minimizes the prediction error between the measured output of the true closed-loop system and the predicted closed-loop output. 
The use of these methods requires the knowledge of the controller.
Considering the case where the input can be delayed by $d$ samples, the plant is described by
\begin{equation*}
G(q^{-1})=\frac{q^{-d}B(q^{-1})}{A(q^{-1})},
\end{equation*}
where $A$ and $B$ are defined as in \eqref{eq:TFbase}.
The plant is operated in closed-loop with an RST digital controller (without lack of generality). 
Introducing the polynomials $A^*(q^{-1})$ and $B^*(q^{-1})$ such that $A(q^{-1})= 1+q^{-1}A^{*}(q^{-1})$ and $B(q^{-1})= q^{-1}B^{*}(q^{-1})$, the output of the plant operating in closed-loop is (see Figure~\ref{fig8_2a}):
\begin{eqnarray*}
y(t+1)&=&-A^*y(t)+B^*u(t-d)+A\eta(t+1)\\
&=&\theta^T \varphi_d(t)+A\eta(t+1),
\end{eqnarray*}
where $u(t)$ is the plant input, $y(t)$ is the plant output, $\eta(t)$ is the output noise, $\theta$ is defined in \eqref{eq:theta} and:
\begin{eqnarray*}
\varphi_d^T(t)&=&[-y(t)\ldots, -y(t-n_a+1), \\
&& u(t-d)\ldots,u(t-n_b+1-d)] ,\\
u(t)&=&-\frac{R}{S}y(t)+r_u,
\end{eqnarray*}
where $r_u$ is the external excitation added to the output of the controller (input of the plant) for identification purpose. 

The  \textit{a priori} predictor of the closed-loop can be expressed as:
\begin{equation*}
\hat{y}^{\circ}(t+1)=-\hat{A}^{\ast}(t) \hat{y}(t)+\hat{B}^{\ast}(t) \hat{u}(t-d)=
\hat{\theta}^T(t) \phi_d(t), 
\end{equation*}
where $\hat{\theta}$ is defined in \eqref{eq:Estpar} and
\begin{eqnarray}
\phi_d^T(t)&=&[-\hat{y}(t) \ldots, -\hat{y}(t-n_a+1),\nonumber\\
&& \hat{u}(t-d)\ldots, \hat{u}(t-n_b+1-d)], \label{eq_uBF_CLOE} \\
\hat{u}(t)&=&-\frac{R}{S} \hat{y}(t)+r_u.
\end{eqnarray}
The a posteriori predictor of the closed loop can be expressed as
\begin{equation*}
\hat{y}(t+1)=\hat{\theta}^T(t+1) \phi_d(t). \label{918A}
\end{equation*}
The  \textit{a priori} closed-loop prediction (output) error is
\begin{equation}
\varepsilon^{\circ}_{CL}(t+1)=y(t+1)-\hat{y}^{\circ}(t+1),
\end{equation}
and the {\em a posteriori} closed-loop prediction error is
\begin{equation}
\varepsilon_{CL}(t+1)=y(t+1)-\hat{y}(t+1).
\end{equation}
The algorithm of \eqref{eq_3_2_51}-\eqref{eq_3_2_55} is used for closed-loop identification, but with $\varepsilon$ replaced by $\varepsilon_{CL}$ (a priori and a posteriori) and with the regressor vector $\phi_d(t)$ given by \eqref{eq_uBF_CLOE} instead of \eqref{eq:phi}. This is called the \textit{closed-loop output error} (CLOE) algorithm.

\subsection{Iterative re-design of the controller}\label{sec:iterative}

As indicated at the beginning of this section, in the technique of iterative closed-loop identification and controller re-design, after an identification in closed loop during a certain time horizon in the presence of an external excitation, the estimated plant parameters are used to re-design the controller.
The control design and the desired performance are exactly the same as those used for the case when an open loop identified model has been used. 
A criterion of performance evaluation has to be defined, and the procedure can be stopped if the performance is no more improved after a number of iterations. 
Reported experimental results (see \cite{Landau2006,Landau11}) indicate that the first two iterations are those that lead to the most significant performance improvement.

\subsection{Experimental results for iterative identification in closed loop and controller re-design}

The iterative identification in closed loop and controller re-design method is implemented on the throttle valve as follows:
\begin{enumerate}
	\item the initial RST controller design is done according to the robust method described in the \emph{Robust control design} sidebar using a generic (known \emph{a priori}) set of parameters $\theta$;
	\item a persistently exciting signal (PRBS) is applied (added to the reference) during a specific period of time and the vector of system parameters $\hat{\theta}$ is evaluated using the CLOE algorithm;
	\item the tunning of the robust RST controller is updated using the estimated vector of parameters $\hat{\theta}$ and implemented in the closed loop;
	\item the previous two steps are repeated in a loop.
\end{enumerate}
\begin{figure}[ht]
	\begin{center}
		\includegraphics[width=\columnwidth]{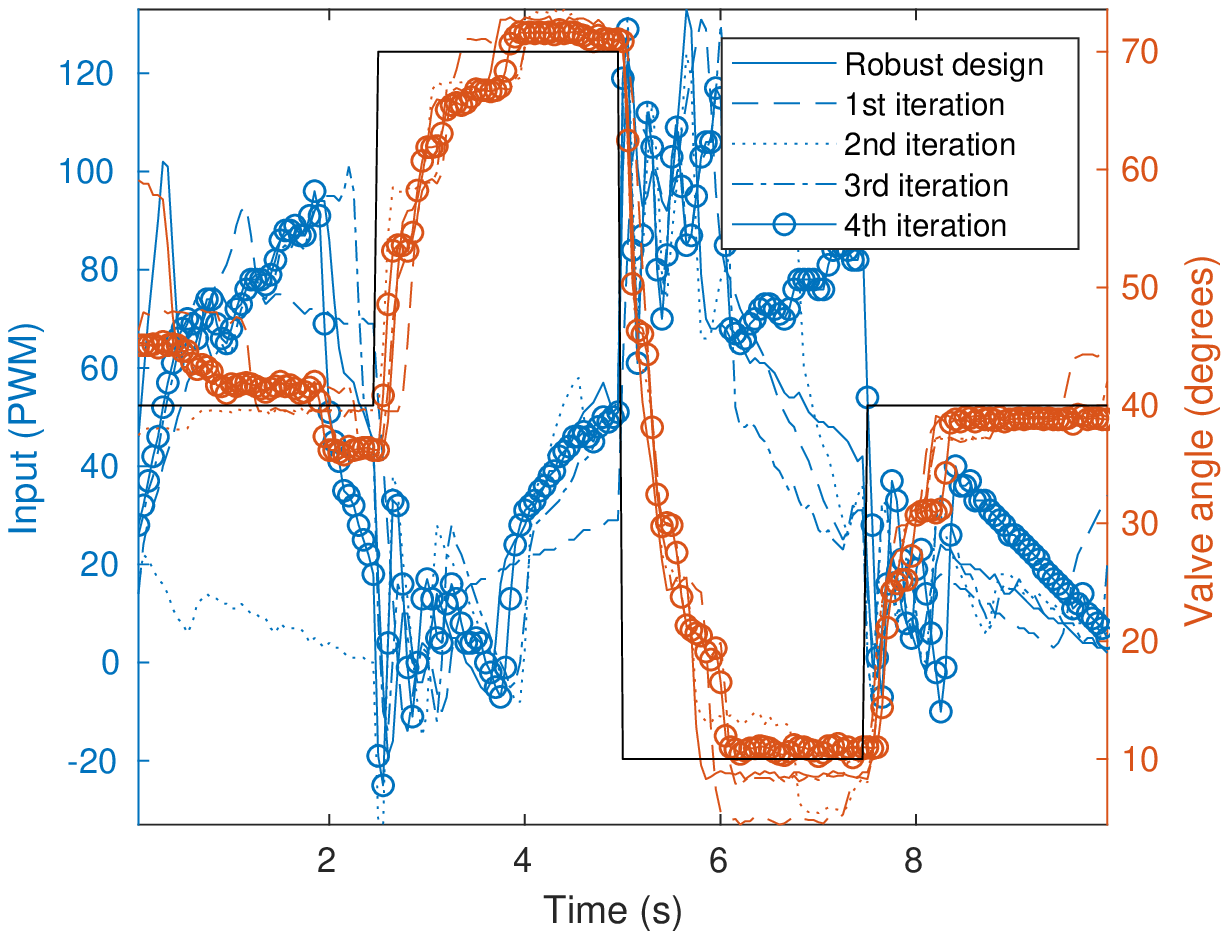}
		\caption{Performance evaluation of the adaptive method \emph{iterative identification in closed loop and controller re-design} on \emph{Experiment 6}.
	Left axis: PWM inputs in blue, right axis: closed-loop responses in orange and tracked reference in black. The response corresponding to the \emph{Robust control design} sidebar (plain line) is compared with successive iterations of the adaptive design (other types of lines). A more accurate tracking is obtained with the adaptive method after a few iterations.
	}
		\label{Fig:Adaptive_exp}
	\end{center}
\end{figure}
The resulting algorithm is developed with {Arduino IDE}\textsuperscript{\tiny\textregistered} and uploaded in the Arduino Mega 2560\textsuperscript{\tiny\textregistered} board.
Comparing step by step the different algorithm steps with the same calculations done with Matlab\textsuperscript{\tiny\textregistered}, noticeable numerical discrepancies are noticed.
This could be expected from the matrix operations / inversions involved in the design and may be of particular interest to illustrate the problems of floating-point and real-time computations in an embedded environment.
A more robust parameter estimation is obtained by increasing the power spectrum of the PRBS at lower frequencies (setting the duration of the longest pulse of the PRBS to 1.6\,s instead of 0.8\,s).
The excitation signal is set to 300 samples, which does not include the full PRBS but gives a satisfactory parameter convergence.
Each sequence of parameter estimation and controller update is followed by an evaluation of the tracking efficiency, with the scenario depicted on Figure~\ref{Fig:Adaptive_exp}.
Comparing the tracking efficiency for different iterations on  Figure~\ref{Fig:Adaptive_exp}, we notice consistent closed-loop responses and more noticeable improvement (compared to the robust PI design) at the 4$^{th}$ iteration.


\begin{sidebar}{Control education perspective}

The content of this paper is the backbone of two series of labs: one for a class on \emph{Modeling and System Identification} 
and one for a class on \emph{Adaptive Control}.

The labs on modeling and system identification consist of 9h of classes (divided in two sessions).
The class concepts of experiment design for system identification (nonlinear dynamics, preliminary experiments and choice of the sampling interval, informative experiments, PRBS input), non-parametric identification (transient response, spectral analysis), and parameter estimation in linear models (basic principles of parameter estimation, minimizing prediction errors, linear regressions and least squares, choice of the model structure) are illustrated.
The use of Arduino IDE\textsuperscript{\tiny\textregistered} to program the PRBS using shift registers is requested as a personal work.
While controller design is not part of the class, a preliminary work on the digital implementation of a PI controller is asked to the students, illustrating the methods available to discretize continuous dynamics.
It is also useful to remind the students that despite the simplicity of the identified model compared to the system nonlinear dynamics, it is perfectly suitable for control objectives.
Extra optional questions comparing different controller designs are proposed, for the students who finish the lab ahead of time.
The valve topic is also used as a motivating example (in the form of a homework) for a lesson on bond graphs, thus making the students familiar with the physical modeling aspect prior to the class.

The labs for adaptive control consist of 12h of classes (divided in 4 sessions).
This sequence of labs aim to illustrate the concepts of parameter adaptation algorithms, closed-loop system identification, controller redesign, and robust control design and parameter estimators for adaptive control.
The preliminary work focuses on RST control design for a first order discrete model, first assigning only two dominant poles for the closed loop and then assigning also an additional auxiliary pole, taking into account the robustness issues.
This implies to have a clear idea on the available degrees of freedom of the controller to achieve a given objective.
The Arduino\textsuperscript{\tiny\textregistered} implementation is also requested.
The second part of the lab is on parameter estimation and parameter adaptation algorithms in linear models.
A preliminary Matlab\textsuperscript{\tiny\textregistered} code is given, and the students have to analyze the structure of the algorithm and associate it with the equations derived in class.
The convergence of the parameters and of the adaptation gain have to be visualized and understood, first with Matlab\textsuperscript{\tiny\textregistered} and then on the experimental test bench.
The last part of the lab is on the adaptive design of digital controllers.
The RST design and the parameters adaptation algorithm are combined to implement the identification in closed loop and controller re-design method.
If there is enough time, the students can investigate more advanced topics such as model reference adaptive control, for tracking and regulation with independent objectives.
\end{sidebar}

\section{Conclusions} 

In this work, we structure the key ingredients for learning data-driven control with the experimental test bench of a throttle valve.
We first focus on understanding the dynamics of the valves, with specific investigations of the nonlinear behavior, of the transient response in the frequency domain, and of the adequacy of a simple PI controller ensuring the desired performance.
Despite the observed complexity of the dynamics, it is shown that a linear design in the discrete-time framework offers a valid solution to the control problem.
A more complex design that also takes into account a robustness objective is also presented.
The evaluation of the sensitivity functions allows assessing the robustness of the design and designing the digital controller accordingly.
The last control design involves an adaptive method in which the controller learns the process parameters online and updates the feedback gains when new measurements are received.
Despite the number of calculations associated with the real-time operation of the adaptive controller, it is successfully embedded into the experimental test bench.
The organisation of the labs organized at Universit\'e Grenoble Alpes to teach this material is presented.

A first perspective with the use of this experimental test bench for teaching purposes is to implement and compare other adaptive methods, considering that the controller can adapt to different operating points.
With the spreading use of Python\textsuperscript{\tiny\textregistered} and its recent inclusion in the teaching programs, another interesting perspective is to design feedback control methods using this software and investigate the use of machine learning toolboxes in this framework.

\bibliography{bib_throttle2}

\end{document}